\documentclass[amssymb,amsmath,aps]{revtex4}
\usepackage{natbib, graphicx, subfigure, subeqnarray,fancyhdr,amsmath,epstopdf,multirow,color, mathrsfs, amssymb, cancel}
\usepackage{natbib,graphicx,subfigure,subeqnarray,fancyhdr,amsmath,epstopdf,multirow,color, mathrsfs, amssymb}
\usepackage{graphicx,subfigure,subeqnarray,fancyhdr,amsmath,epstopdf,multirow,color, mathrsfs,amsmath,amssymb, natbib}
\newcommand{\y}{\hat{\mathbf{y}}}
\newcommand{\x}{\hat{\mathbf{x}}}
\newcommand{\z}{\hat{\mathbf{z}}}
\newcommand{\uvec}[1]{\mathbf{\hat{{#1}}}}
\newcommand{\I}{\boldsymbol{\rm{I}}}
\newcommand{\pe}{_{\perp}}
\newcommand{\pa}{_{\parallel}}
\newcommand{\sh}{\dot{\boldsymbol{\gamma}}}

\newcommand{\intarc}{\int_0^{\Lambda}}

\def\ub{{\bf u}}

\begin{document}
\title{An empirical resistive-force theory for slender biological filaments in shear-thinning fluids}
\author{Emily E. Riley and Eric Lauga}
\email{e.lauga@damtp.cam.ac.uk}
\affiliation{Department of Applied Mathematics and Theoretical Physics, University of Cambridge, CB3 0WA, United Kingdom}
\date{\today}

\begin{abstract}

Many cells exploit the bending or rotation of flagellar filaments in order to self-propel in viscous fluids.  While appropriate theoretical modelling is available to capture flagella locomotion in simple, Newtonian fluids, formidable computations are required to address theoretically their locomotion in complex, nonlinear fluids, e.g.~mucus.  Based on experimental measurements for the motion of rigid rods in non-Newtonian fluids and on the classical Carreau fluid model,  we propose  empirical extensions of the classical Newtonian resistive-force theory to model  the waving of slender filaments in  non-Newtonian fluids. By assuming the flow near the flagellum to be locally Newtonian, we propose a self-consistent way to estimate the typical shear-rate in the fluid, which we then use to construct correction factors to the Newtonian local drag coefficients. The resulting non-Newtonian resistive-force theory, while empirical, is   consistent with the Newtonian limit,  and with the experiments. We then use our  models to address waving locomotion in non-Newtonian fluids, and show that the resulting  swimming speeds are systematically lowered -- a result which we are able to capture asymptotically and to interpret physically. An application of the models to recent experimental results on the locomotion of  {\it Caenorhabditis elegans} in polymeric solutions shows reasonable agreement and thus captures the main physics of swimming in shear-thinning fluids. 

\bigskip

\end{abstract}
\maketitle
\section{Introduction}

Biological locomotion is crucial on microscopic scales. From finding new food sources and evading predators for individual swimmers, to fertilisation and flow induction in higher  organisms, 
the generation of fluid flows at micro-scale is an important field of study \cite{vogel96}, dating all the way back to the advent of the microscope and the first observations of bacteria and spermatozoa by Leeuwenhoek \cite{booklittleanimals}. 

Below the millimetre scale,  viscous effects tend to dominate and macro-scale methods of propulsion that rely on inertial momentum transfer are often ineffective.  In order to produce  continuous motion at low Reynolds number, a swimmer must move its body periodically but avoid time-reversible motion \cite{Purcell1977}. For swimmers with back-and-forth motion, the viscous drag during the power stroke must be greater than that during the recovery stoke, and  it is this drag anisotropy that leads to locomotion.  Thin rodlike appendages, usually termed flagella, are used by both eukaryotic \cite{Ginger2008} and prokaryotic \cite{Jarrell2008} cells, as well as man-made swimmers \cite{BioroboticsBook}, in order to induce   propulsion in the absence of inertia.    Due to the difference in drag coefficients for rods moving in a fluid perpendicular vs.~parallel to their long axis,  the  time reversibility is broken for travelling wave-like motion of  the appendages of swimmers, which allows the generation of propulsion   \cite{Cohen2010}.  Although flexible planar wave motion of eukaryotes and rigid rotation of helical prokaryotic flagella evolved separately billions of years ago, they both take advantage of this anisotropy in drag via the large aspect ratio of their flagella \cite{BookSmall}.

In order to describe swimming induced by long, slender flagella,  resistive-force theory was proposed over 60 years ago \cite{Hancock1953,Gray1955} and has subsequently been improved upon \cite{LighthillLecture}. 
The basic idea is to approximate the perturbation induced by the flagellum on the fluid as a line of flow singularities (point forces). For a radius of curvature of the flagellar waveform  much larger than the diameter of the flagellum then at leading-order in the aspect ratio of the flagellum, the local velocity linearly determines the local force density  on the flagellum. The drag can then be decomposed into the perpendicular and parallel components in this local region \cite{Blum1973,Powers2009}. Corrections to resistive-force theory have been made to improve its accuracy by increasing the number of terms in the expansion thereby including analytically hydrodynamic interactions between different portions of the flagellum in a systematic fashion \cite{Batchelor1970, Cox1970}. 
Further refinements include  slender-body theory \cite{Johnson1980,LighthillLecture} which  provides greater accuracy leading to better qualitative and quantitative approximations \cite{Rodenborn2013}, compared to resistive-force theory but typically requires numerical evaluation. 

While it is only valid asymptotically, and is only logarithmically accurate,  resistive-force theory has been  shown to be a good approximation in many instances \cite{Johnson1979}. Without resistive-force theory the only analytical insight into low-Reynolds number swimming would come  from small-amplitude swimmers \cite{Taylor1951,taylor52}. 
In contrast, resistive-force theory allows calculations  on the kinematics of swimming at  large amplitudes, as relevant  to real flagellar motion.  For this reason it provides a good intermediary model between simple small-amplitude results and complex computations and has  had great success in describing the locomotion of microorganisms in Newtonian fluids \cite{Yu2006,Friedrich2010,Berman2013}.

Beyond Newtonian fluids, many biologically-relevant fluid environments have shear-dependent viscosities, including for example  lung mucus, cervical mammalian mucus \cite{velezcordero13} and soil \cite{SoilThesis}. Prompted by the success of resistive-force theory and the biological relevance of non-Newtonian fluids, it is natural to ask if it would be possible to  derive a resistive-force theory for nonlinear fluids. 

Most  theoretical studies of motion in shear-thinning fluids focus on small-amplitude asymptotics. 
This includes small-amplitude perturbations of Taylor's swimming sheet \cite{velezcordero13}  and squirming motion on a spherical surface \cite{Datt2015}.  Of course,  real biological  swimmers fall beyond the asymptotic small-amplitude limit.  In order to probe theoretically large-amplitude motion in non-Newtonian fluids, complex numerical simulations are required,  such as those performed on finite high amplitude sheets \cite{Shelley2013,li15} and the thick nematode \emph{Caenorhabditis elegans} \cite{Becca2014} in viscoelastic fluids, and those performed on a variety of swimmer types in shear-thinning fluids \cite{Loghin2013}.  
Though they provide important novel physical  insight, such computational approaches  are, by nature,  difficult to extrapolate to other geometries, waving kinematics etc. It would thus be  useful to have a modelling tool  easily implementable and allows us to tackle a variety of flagellar kinematics. 

 The study of bodies moving in non-Newtonian fluids has a long and rich history.  Dating back over a century, the earliest studies of single particles in Newtonian Stokes flow found analytic solutions to axisymmetric particles, such as spheres \cite{Stokes1851}, prolate and oblate ellipsoids, lenses and spherical caps \cite{Bowen1973}.
However for extension to cylinders and other thin shapes such as double headed cones, full steady state solutions cannot be found due to the Stokes paradox in two-dimensions  and approximate solutions can only be obtained for thin cylinders \cite{BookHappelBrenner}.  Analytical studies on the motion of rigid spheres in non-Newtonian fluids  have been conducted using a variety of shear-thinning  models including power-law \cite{BookDrops}, Carreau \cite{Rodrigue1996} and Ellis \cite{Hopke1970}.  Although exact solutions can be found, the results reported from power-law fluid models often do not agree with one another and with experimental results \cite{BookDrops}. Greater success and  agreement with  experiments \cite{BookDrops}  has been obtained with the Carreau and Ellis fluid models.  Analytical studies in this case used  expansion of small non-Newtonian effects for the Carreau fluid and extremum principles for the Ellis model   \cite{Rodrigue1996, Hopke1970}  while  numerical approaches required fitting  external parameters to the data (Carreau, \cite{Bush1984}).

In non-Newtonian fluid mechanics,  empirical fitting is a key modelling approach. Due to the nonlinear nature of the fluid,  parameters are often fit to certain shear-thinning indices, other rheological properties, or shape parameters,  allowing prediction of behaviour in a variety of  shear-thinning fluids. As studies branch away from rigid spheres in infinite fluids, the complexity of calculations increases again due to orientation considerations, and past theoretical studies mostly rely on numerics. While it was shown analytically that  the Stokes paradox vanishes for cylinders in power-law fluids \cite{Tanner1993}, attempting to extend this to Carreau fluids has proved problematic.  Furthermore, the majority of the work on rods and  cylinders in non-Newtonian fluids has focused on small but finite Reynolds numbers, motivated by industrial applications.  In this regime the shear-thinning effects of the fluid tend to be  negligible at very small Reynolds numbers   or close to walls \cite{Hsu2005}.  In the creeping flow limit the drag coefficients calculated numerically show reduction compared to those calculated in a Newtonian fluid \cite{Hsu2005, Machac2002}, similar to experimental results \cite{Chhabra2001}, where both motion  of  cylinders orientated parallel and perpendicular to their motion  was investigated.  The role of the aspect ratio of cylindrical rods has been studied in inelastic and elastic fluids, showing that  drag coefficients reduce  by about one  order of magnitude over aspect ratios ranging from 1/150 to 1/10 \cite{Cho1992}. Comparing these experimental result to semi-empirical predictions, despite qualitative agreement, the drag coefficient was overestimated by theoretical predictions \cite{Park1988, Manero1987}. Comparable studies have also probed wall effects \cite{Hsu2005, Unni1990}, different  cross sections \cite{Lee1986} and interactions between particles \cite{Cordero2011}.

  Returning to the impact on biological swimmers, the fundamental physical problem to tackle concerns the force generation by beating flagella.   Physically, we expect that flagella waving in shear-thinning fluids will experience two important effects \cite{gomez16}. One is a local influence due to changes in the viscosity. If a body is subject to a Stokes-like force law and the viscosity of the fluid decreases, then the local force will decrease \cite{Chhabra2001}, and swimmers will then experience either enhanced or decreased locomotion based on the detailed balance between drag and thrust. The second effect, more subtle, is nonlocal and due to the change in the flow field around the body. Bodies moving in shear-thinning fluids are expected to be surrounded by low-viscosity regions, themselves embedded in high-viscosity domains. This  thus makes swimming in a shear-thinning fluid akin to swimming under (soft) confinement, which might lead to an increase  of propulsion \cite{Man2015,li15}. In this paper we propose a theoretical  model for swimming in shear-thinning fluids addressing the first, local, effect by building an empirical extension of resistive-force theory in complex fluids. Specifically,  and similarly to recent work in granular media \cite{Zhang2014},  we propose to use experimental results on rods falling in shear-thinning fluids to obtain estimates on the drag coefficients acting along slender bodies (\S \ref{building}). We then quantify the impact of these coefficients on waving locomotion (\S \ref{Splanar}) and compare our predictions with recent experiments on {\it C. elegans} (\S \ref{Snematode}).

\section{Building a non-Newtonian resistive  force theory}
\label{building}
\subsection{Methodology}
The aim of this paper is to propose a new, nonlinear relationship  between the velocity of a slender filament relative to a background fluid flow and the local hydrodynamic force density acting on it. We should point out at the outset that we are not deriving a rigorous mathematical model from first principles, as this is in fact virtually impossible due to the  nonlinearity of the constitutive relationships, but instead seek to describe filament motion in shear-thinning fluids empirically.  

Two  approaches are used to calculate the non-Newtonian drag  coefficients. The first one is an empirical fit to experimental measurements of sedimenting rods in shear-thinning fluids, and thus is directly built from experimental data.  The second approach  is an ad-hoc model based on the Carreau viscosity-shear-rate relationship.  Since in shear-thinning fluids the shear viscosity of the fluid is a function of the shear-rate, we first need, in this case, a method to estimate accurately  the local shear-rate around the moving filament. We do so by approximating the flow as locally Newtonian, allowing us to exploit elementary  flow calculations. With this local, instantaneous value of the shear-rate, we can then incorporate the shear-thinning nature of the fluid though a correction to  the Newtonian drag coefficients and  therefore a nonlinear velocity-force relationship.   For both approaches, we ensure that our methodology is consistent with the Newtonian limit and we carefully examine the limit in which we expect this approach to be valid. 

\subsection{Shear-rate around slender filaments}
\label{Sshear}
\begin{figure}[!]
\begin{center}
 \includegraphics[width=0.49\textwidth]{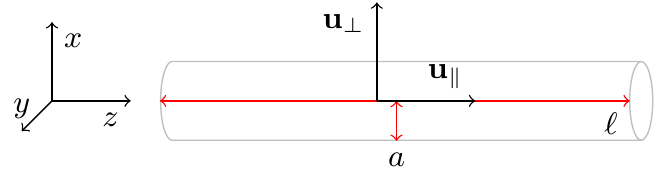}
 \caption{A straight filament of length $2\ell$ and cross-sectional radius $a$ may translate in a fluid along its length (velocity $\ub_\parallel$) or perpendicular to it ($\ub_\perp$). The axis of the cylinder is along the $\z$ direction while $\x$ and $\y$ are in the cross section.}
 \label{figcylinder}
 \end{center}
\end{figure}

\subsubsection{Newtonian flows near filaments}
\label{Srods}

In order to estimate the shear-rates around moving filaments we make a locally-Newtonian assumption.   To describe the drag on a slender filament we decompose the motion along  the directions parallel and perpendicular to the local axis of the filament, leading to two drag
coefficients. For simplicity consider a straight filament of cross-sectional radius $a$ and length $2\ell$ (Fig.~\ref{figcylinder}). 
Following Lighthill's classical analysis \cite{LighthillLecture},  at leading order in the aspect ratio of the filament, the flow near the filament  is described by a local, uniform line distribution of point forces and (potential) source dipoles along the  centreline of the rod. 

When the filament is translating  along its symmetry axis, $\z$, the flow around the cylinder is given by a  linear distribution of point forces, with no need for dipoles, giving a relationship between the velocity field close to the filament, $\ub_\parallel(x,y,z)$, and  the force per unit length acting along the filament  $f\pa\z$,  as
\cite{LighthillLecture}
\begin{equation}\label{upara}
\mathbf{u}_{\parallel}(x,y,z) = \frac{f\pa}{4\pi\eta_0}
\left[\ln\left(\frac{4\ell^2}{r^2}\right)-1\right]\z,
\end{equation}
where $r=\sqrt{x^2+y^2}$ is the distance from the axis of the filament and $\eta_0$ is the Newtonian viscosity.  At the surface of the cylinder $r=a$, the parallel drag coefficient, $b\pa$,  is given by 
\begin{equation}
\frac{f\pa}{{u}_{\parallel}|_{r=a} } \equiv b\pa = \frac{4\pi\eta_0}{\ln\left({4\ell^2}/{a^2}\right)-1}.
\end{equation}

When the filament is translating  perpendicular to its axis (here, the $\x$ direction) then  a combination of point forces and source dipoles are required to model the flow  \cite{LighthillLecture}. The velocity field near the filament  is now given by 
\begin{equation}\label{uperpr}
\mathbf{u}_{\perp}(x,y,z) = \frac{f\pe}{8\pi\eta_0}\left(
\begin{array}{c}\displaystyle
\ln\left(\frac{4\ell^2}{r^2}\right)+\frac{a^2}{r^2}+\frac{2x^2}{r^2}\left(1-\frac{a^2}{r^2}\right)\\
\displaystyle\frac{2xy}{r^2}\left(1-\frac{a^2}{r^2}\right)\\
0
\end{array}
\right),
\end{equation}
in the $(\x, \y, \z)$ coordinate frame where  $f\pe$ is the force  acting on the filament per unit length. On the surface of the filament we have then 
\begin{equation}\label{uperp}
\mathbf{u}_{\perp}|_{r=a} = \frac{f\pe}{8\pi\eta_0}\left(
\begin{array}{c}
\displaystyle\ln\left(\frac{4\ell^2}{a^2}\right)+1\\
0\\
0
\end{array}
\right),
\end{equation}
\emph{i.e.}~the filament translates in the $\x$ direction and the velocity is uniform around its surface. Similarly  to the  motion parallel to the filament axis the perpendicular drag coefficient, $b\pe$, is given by
\begin{equation}
\frac{f\pe}{{u}_{\perp}|_{r=a} } \equiv  b\pe =  \frac{8\pi\eta_0}{\ln\left({4\ell^2}/{a^2}\right)+1}.
\end{equation}

In order to  extend these  drag coefficients obtained for a straight filament to smooth curved filaments   a relevant length $\ell$ over which the filament can be approximated as straight is required.  In the application of resistive-force theory to  travelling waves along sperm flagella, Gray and Hancock chose for $\ell$ the wavelength $\lambda$ as the only relevant length scale along the swimmer but without mathematical justification \cite{Gray1955}. In subsequent work, Lighthill showed mathematically by considering a periodic distribution of flow singularities that  $\ell  \approx 0.09\lambda$ was the relevant length scale along a periodic wave of wavelength $\lambda$ \cite{LighthillLecture}.  This is the choice we make here to address waving motion with the understanding that other filament kinematics might require a different choice.

\subsubsection{Shear-rates}

In order to propose  drag coefficients to use with the Carreau model (or any other shear-thinning empirical fluid model \cite{Morrisonbook}) we require knowledge of the shear-rates in the fluid near the filament.  In order to estimate  shear-rates we again use  Lighthill's calculations  \cite{LighthillLecture}. 

In case of parallel motion, we calculate the velocity gradient, $\nabla\mathbf{u}$, using Eq.~\eqref{upara}. In cylindrical polar co-ordinates $(\hat{\mathbf{r}}, \hat{\mathbf{\phi}}, \z)$ the only non-zero term is given by
\begin{equation}
\frac{\partial u_{z_{\parallel}}}{\partial r} = -\frac{f\pa}{2r\pi\eta_0},
\end{equation}
and thus the shear-rate tensor, $\sh_{\parallel} $, is given by
\begin{equation}
\sh_{\parallel}=
-\frac{f\pa}{2r\pi\eta_0}
\left(\begin{array}{ccc}
0 & 0 & 1\\
0 & 0 & 0\\
1 & 0 & 0
\end{array}
\right).
\end{equation}
For a filament moving locally in the   direction perpendicular to its axis, the shear-rate tensor, $\sh_{\perp}$,  is obtained by taking the gradient of the flow in  Eq.~\eqref{uperpr}, such that
\begin{equation}
\sh_{\perp} = \frac{f\pe}{2r\pi\eta_0}\left(
\begin{array}{ccc}
\cos\phi(-1+a^2/r^2) & \frac{a^2}{r^2}\sin\phi & 0\\
\frac{a^2}{r^2}\sin\phi & \cos\phi(1+a^2/r^2) & 0\\
0 & 0 & 0
\end{array}
\right).
\end{equation}

In order to capture  how the viscosity changes near the filament, the total shear-rate near the filament is required. Indeed, a change in the viscosity  due, for example, to a perpendicular motion will then affect the apparent viscosity for movement in the parallel direction, and vice versa.  In other words, when the fluid is not Newtonian  we can no longer consider perpendicular and parallel motions separately but need to include both solutions together.   To find the total shear-rate, $\sh_{tot}$,  we exploit  linearity and add the perpendicular and parallel solutions together to find on the filament $r=a$, the tensor
\begin{equation}
\sh_{tot} = \frac{1}{2a\pi\eta_0}\left(
\begin{array}{ccc}
0& f_{\perp}\sin\phi & -f\pa\\
f_{\perp}\sin\phi & 0& 0\\
-f\pa & 0 & 0
\end{array}
\right).
\end{equation}
The first shear-rate invariant is zero at $r=a$, thus we calculate the second shear-rate invariant 
$|\sh|^2= {\rm tr} (\sh^2)/2$ where $\rm tr$ refers to the trace of the tensor \cite{Morrisonbook}, such that
\begin{equation}
 |\sh|^2_{tot} = \frac{\sin^2\phi f_{\perp}^2+f_{\parallel}^2}{(2a\pi\eta_0)^2}\cdot
\end{equation}
To find the average value of the shear-rate invariant around the surface of the cylinder we integrate around the cylinder axis ($\phi$) and divide by $2\pi$, and define the average shear-rate on the surface due to both perpendicular and parallel motion of the rod as 
\begin{equation}\label{eq=shtot}
\dot{\gamma}_{avg} = \frac{\sqrt{f_{\perp}^2+2f_{\parallel}^2}}{2\sqrt{2}a\pi\eta_0}\cdot
\end{equation}
Note that beyond this local flow, hydrodynamic singularities far from the local portion of the filament also contribute to flow and shear-rates, but these will be at least $O(a/\ell)$ smaller, and are thus sub-dominant \cite{LighthillLecture}. In the slender limit, the result in Eq.~\eqref{eq=shtot} gives therefore the leading-order value of the mean square shear-rate along the filament.

Finally,  Eq.~\eqref{eq=shtot} relates the local shear-rate to the local force density. In order to be used in a resistive-force theory-type approach, we need instead to have a relationship between the shear-rate and the local velocity.  To quantify the forces on the filament in a self-consistent fashion we write
\begin{eqnarray}
f_{\perp}=b_{\perp}u_{\perp}, &\rm{and}& f_{\parallel}=b_{\parallel}u_{\parallel},
\end{eqnarray}
and thus the average shear-rate around the rod is given by 
\begin{equation}\label{eq=shavg}
 \dot{\gamma}_{avg} = \frac{\sqrt{b_{\perp}^2u_{\perp}^2+2b_{\parallel}^2u_{\parallel}^2}}{2\sqrt{2}a\pi\eta_0}\cdot
\end{equation}
For a given velocity, we thus obtain that the locally-Newtonian assumption leads to a shear-rate independent of the viscosity, since both drag coefficients scale linearly with the viscosity (\emph{i.e.}~we get $\sh\sim u/a$). 
We simplify the shear-rate by defining the shear-rate velocity as
\begin{equation}
 u_{\sh} = \sqrt{u_{\perp}^2+2\frac{b\pa^2}{b\pe^2}u_{\parallel}^2},
\end{equation}
such that
\begin{equation}\label{shear_final}
 \dot{\gamma}_{avg} = \frac{b_{\perp}u_{\sh}}{2\sqrt{2}a\pi\eta_0}\cdot
\end{equation}

\subsection{Notation}

Depending on the model, shear-thinning fluids may be  characterised by a number of rheological parameters. For example, for a Carreau-like fluid or a power-law-like fluid \cite{Morrisonbook}, one rheological parameter is the  shear-thinning index, $0<n<1$, that describes by how much the viscosity reduces with increasing shear-rate ($n=1$ being the Newtonian limit). A Carreau-like fluid is also characterised by the  critical shear-rate, $1/\Gamma$,  at which the fluid transitions from a Newtonian fluid, with viscosity $\eta_0$, to a shear-thinning fluid.  In order to describe swimming through a non-Newtonian fluid, the Newtonian drag coefficients $(b\pa, b\pe)$
 are replaced by their non-Newtonian counterparts $(b_{NN\pa}, b_{NN\pe})$.  
To quantify the non-Newtonian drag coefficients we introduce two correction factors, $( R\pa, R\pe)$, defined as
\begin{equation}\label{eq=R}
 R\pa= \frac{b_{NN\pa}}{b\pa},
\end{equation}
and,
\begin{equation}
  R\pe  = \frac{b_{NN\pe}}{b\pe}.
\end{equation}
If these drag coefficients are to describe motion in a shear-thinning fluid then they are likely to depend on the local shear-rate (and thus both local velocity and the Newtonian drag coefficients) in a nonlinear fashion, as well as on all the rheological parameters of the fluid and the geometrical parameters of the filament. We propose two empirical approaches in this paper, one based on experimental results and one based on the ad-hoc Carreau model. In order to  distinguish between the correction factors in our two models below we use the subscript  $E$ to denote the correction factor derived from experiments while the subscript $C$ will be used to denote the Carreau  correction factor.

\subsection{A non-Newtonian resistive-force theory from empirical data}
We build our first empirical resistive-force theory from the experimental  results of Ref.~\cite{Chhabra2001}. 
In this study, measurements were made of the sedimentation speed of rigid rods  under gravity into a variety of fluids at low Reynolds numbers ($0.01<\mathrm{Re}<0.27$ based on their terminal velocity). The orientation of the rods was either aligned with gravity or  perpendicular to it. 
The forcing from gravity is known and velocities are measured, allowing access to the drag coefficients. 
The rods used are a variety of materials (perspex, polyvinyl chloride, alluminium and stainless steel) with aspect ratios, $\alpha=d/L$, ranging from 1/10 to 1, where $L$ is the rod length and $d$ is the rod diameter.  The non-Newtonian fluids in which the rods are dropped are shear-thinning viscoelastic fluids, with critical times ranging from $0$~s$<\Gamma<19$~s, and shear-thinning indices spanning $0.6<n<1$ \cite{Chhabra2001}. Rheometry data from Ref.~\cite{Chhabra2001} shows that the viscosity vs.~shear-rate relationship for each of the five fluids probed can be described by the Carreau model, however they have non-zero first normal stress differences.

\begin{figure}[!]
 \includegraphics[width=0.49\textwidth]{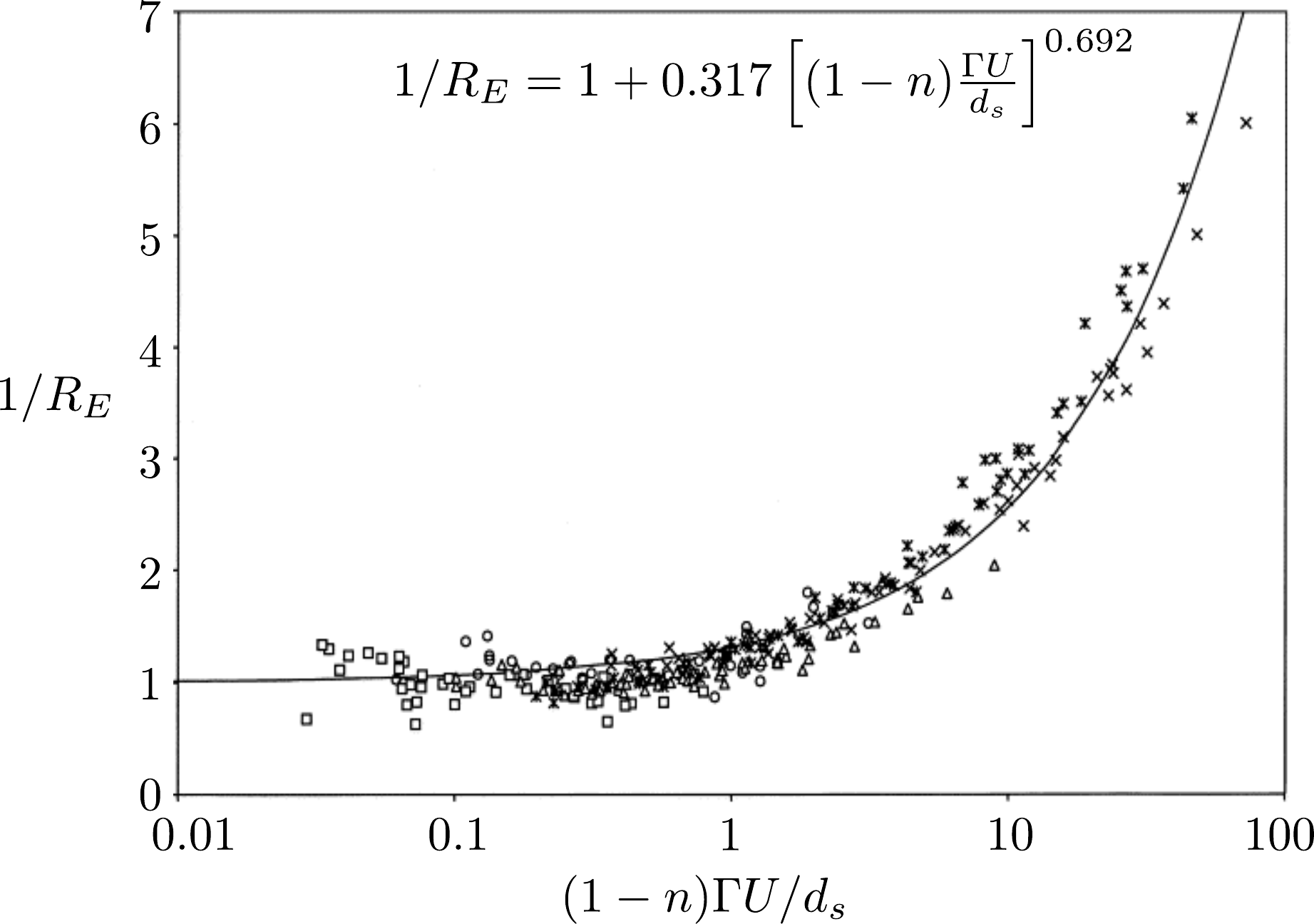}
 \caption{Inverse of the drag coefficient correction factor, $R_E$, as obtained experimentally by measuring the sedimentation speed of  cylindrical rods under gravity  in a variety of shear-thinning fluids \cite{Chhabra2001}. Results from both vertical ($\parallel$) and horizontal ($\perp$) rod orientations are shown and different  fluids are marked by different symbols.  Here $n$ is the power index of the fluid, $\Gamma^{-1}$ the critical shear-rate for transition to shear-thinning behaviour, $U$ the velocity of the rod, and $d_s$ a relevant length scale characterising the rod (see text). Inset: empirical formula, Eq.~\eqref{fit}, proposed to fit all data \cite{Chhabra2001}. 
Adapted and reprinted from Chemical Engineering Science, 56(6), R.P. Chhabra,Kirti Rami,P.H.T. Uhlherr, Drag on cylinders in shear thinning viscoelastic liquids, 2221-2227, Copyright (2001), with permission from Elsevier.}
 \label{fig:chhabra}
\end{figure}

All results from Ref.~\cite{Chhabra2001} are  reproduced in Fig.~\ref{fig:chhabra} where both perpendicular and parallel rod orientations are shown for all five non-Newtonian fluids.  No systematic impact of the orientation of the rod on the experimental results is evident (specifically the change in the sedimentation velocity, i.e.~the correction factor), suggesting therefore that   correction factors are approximately  independent of orientation in these experiments, $R_{E\pe}\approx R_{E\pa}\equiv R_E$.  Based on their data, the authors of Ref.~\cite{Chhabra2001} proceeded to propose an empirical formula fitting their data, namely a correction factor $R_E$  given by 
\begin{equation}\label{fit}
\frac{1}{R_E} =  1+0.317\left[(1-n)\Gamma \frac{|\mathbf{u}|}{d_s}\right]^{0.692},
\end{equation}
where $n$ and $\Gamma$ are as defined earlier, $|\ub|$ is the  magnitude of the rod velocity and  $  d_s = \sqrt[3]{3Ld^2}$ is the equivalent sphere diameter of the rods. The average error between their data point and this empirical best-fitting curve is 12\%.

Based on this, we  use the fit from Eq.~\eqref{fit}  as our first empirical resistive-force theory in non-Newtonian fluids. Specifically we write that the non-Newtonian drag coefficients are given in this case by
\begin{equation}\label{bE}
b_{E_{\perp}} = R_E(|\mathbf{u}|)b_{\perp},\quad b_{E_{\parallel}} = R_E(|\mathbf{u}|)b_{\parallel},
\end{equation}
and we choose $L=\ell$, in-keeping with the calculation of the shear-rate and the derivation of the Newtonian drag coefficients, in the  expression for the effective rod diameter used in Eq.~\eqref{fit}. Importantly, we note that Eq.~\eqref{bE} is fully consistent with Newtonian resistive-force theory in  that for $n=1$, or $\Gamma=0$, we recover the Newtonian solution. 
Furthermore, Eq.~\eqref{bE} is consistent with experimental data shown in Fig.~\ref{fig:chhabra}.

\subsection{A  non-Newtonian resistive-force theory from the Carreau model}
We now attempt to build the second  correction factor from a shear-thinning fluid model.
We assume that the flow is locally Newtonian and thus we can employ the shear-rate from Eq.~\eqref{shear_final} in any shear-thinning fluid model. We  choose  the Carreau model which is  a good fit to many shear-thinning fluids \cite{velezcordero13,Arratia2014} and in particular to the data of Ref.~\cite{Chhabra2001}. In a Carreau model, the  viscosity of the fluid  is given by
\begin{equation}
 \eta = \eta_{\infty}+(\eta_0-\eta_{\infty})\left[1+(\Gamma\dot{\gamma})^2\right]^{\frac{n-1}{2}}.
\end{equation}
The model is well-behaved and shear-thinning for  $0 < n < 1$.  
Here  $\eta_0$ and $\eta_{\infty}$ describe the fluid's Newtonian zero and infinite shear-rate viscosities, respectively. Since high shear-rates are unlikely to be reached, we set $\eta_{\infty}=0$ so that the model simplifies to 
\begin{equation}
\frac{\eta}{\eta_0} = \left[1+\left(\Gamma\dot{\gamma}_{avg}\right)^2\right]^{\frac{n-1}{2}},
\end{equation}
 where $\dot{\gamma} = \dot{\gamma}_{avg}$.
Together with Eq.~\eqref{eq=shavg}, we describe the second correction factor as
\begin{equation}\label{RC}
 R_C = \left[1+\left(\frac{\Gamma b_{\perp}u_{\sh}}{2\sqrt{2}a\pi\eta_0}\right)^2\right]^{\frac{n-1}{2}}.
\end{equation}
As a result, the Carreau non-Newtonian drag coefficients are defined by
\begin{equation}\label{bC}
 b_{C_{\perp}}=R_{C}(u\pe,u\pa)b_{\perp},\quad
  b_{C_{\parallel}}=R_{C}(u\pe, u\pa)b_{\parallel},
\end{equation}
with $R_C$ from Eq.~\eqref{RC}. 
Again, we note the Carreau correction factor is consistent with Newtonian resistive-force theory, and Eq.~\eqref{bE} reduces to unity when $n=1$ or $\Gamma=0$ to recover the Newtonian solution. 
Note that, contrary to the experimental non-Newtonian drag ratio, the Carreau maintains a difference between parallel and perpendicular orientations. 


\subsection{Comparison between the two models}
\begin{figure}[!]
\centering
  \includegraphics[width=0.42\textwidth]{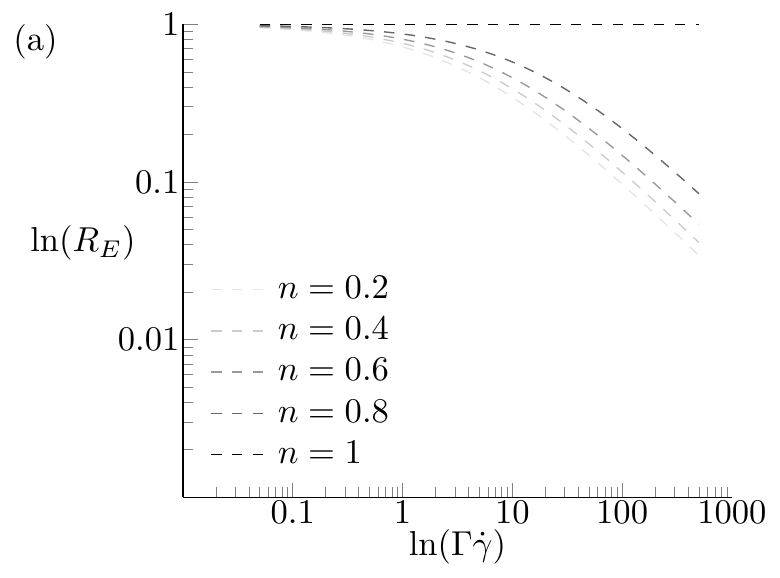}
  \includegraphics[width=0.42\textwidth]{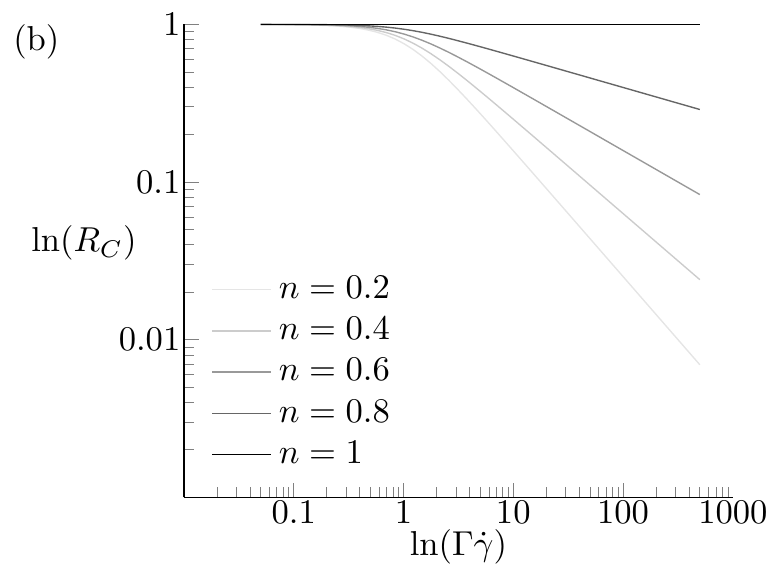}
\caption{Non-Newtonian correction factors  as a function of the dimensionless flow shear-rate for the experimental results ((a) $R_E$) and the Carreau fluid model ((b) $R_C$). 
Increasing shear-thinning indices $n$ are shown from light grey to dark grey with the Newtonian limit $n=1$ shown in black. } 
\label{figeta} 
\end{figure}

We have proposed two methods to estimate non-Newtonian drag coefficients, one based on fitting experimental data (Eq.~\ref{bE}) and one based on using  the classical empirical Carreau  model (Eq.~\ref{bC}). 
A comparison between the non-Newtonian correction factors for these two models is  shown in Fig.~\ref{figeta} where we plot the correction factors as a function of the  local dimensional shear-rate in the fluid  (i.e. $\Gamma \dot\gamma$ for the Carreau model and $\Gamma |\ub|/d_s$ in the  case of the experiments).  Both correction factors show the same qualitative behaviour decreasing with: an increased actuation shear-rate; an increased critical time $\Gamma$; and a reduced shear-thinning index $n$. 
As expected, the nonlinear dependence of $R_C$ on $n$ is stronger than the  linear dependence in the case of $R_E$ and  there is thus a  stronger reduction in drag with smaller values of $n$ in that case.

 Some important differences are however to be expected in the results. The empirical fit described by Ref.~\cite{Chhabra2001} was built from a small range of shear-thinning fluid parameters, thus in order to explore a wider range of $n$ and $\Gamma$ value we will push the experimental model past its true regime of validity. In comparison the Carreau fluid model is valid for all $n$ and $\Gamma$ values. We note also that Carreau fluids have zero first and second normal stress differences, whereas the fluids measure in Ref. \cite{Chhabra2001} have non-zero first normal stress differences which increase with increasing shear rate.

\subsection{Regime of Validity}
\label{regime-valid}
The shear-rate calculated in Section \ref{Sshear} describes the local flow around the filament in the limit where it is asymptotically slender. In that case the relevant shear-rate near the rod is dominated by that induced by the local portion of the filament.   
In order for our local resistive-force theory to be self-consistent,  the fluid viscosity around each  section of the filament must only affected by the movement of said filament section. This requires the shear-rate at the cut-off distance $\ell$ away from the flagellum to be less than the critical shear-rate, $|\sh_C|=1/\Gamma$, at which the fluid becomes shear-thinning. The shear-rate scales as $\sh\sim u_{\sh}/r$, hence the validity of our model is constrained to flows where 
\begin{equation}\label{eq:constraint}
\frac{u_{\sh}}{\ell}\lesssim\frac{1}{\Gamma}\cdot
\end{equation}
For illustration purposes, let us consider the flagellar motion of a spermatozoon with beat frequency $\nu\sim30$~Hz, wavelength $\lambda\sim70~\mu$m \cite{Friedrich2010}, and flagella diameter $2a\sim100$~nm \cite{Powers2009}. For a waving flagella the maximum velocity reached by any rod section is the wavespeed $V=\nu\lambda$, and $\ell=0.09\lambda$. Using $u_{\sh}\approx V$ the constraint in Eq.~\eqref{eq:constraint} simplifies to the inequality $\Gamma\nu\leq0.09$, and for the given swimmer the range of critical times our model can describe is given by $\Gamma\leq3\times10^{-3}$~s. Alternatively, if the fluid properties are given the model is  constrained by a maximum actuation frequency.

\section{Locomotion of waving slender filaments}
\label{Splanar}

In order to illustrate the results given by our non-Newtonian resistive-force theory we apply in this section this modelling approach to study the waving of slender filament as a model for the locomotion of flagellated eukaryotes \cite{LighthillLecture}.

\subsection{Setup}
\begin{figure}[!]
 \includegraphics[width=0.49\textwidth]{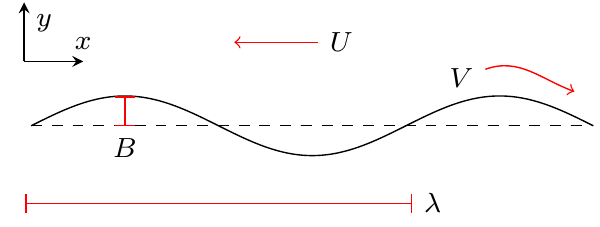}
 \caption{Sinusoidal travelling waveform as a model for the locomotion of flagellated eukaryotic cell. Swimming, with speed $U$, is in the opposite direction to the travelling wave, with wavespeed $V$. See text for notation.}
 \label{wave}
\end{figure}

We consider the swimming of an infinite inextensible filament whose shape deforms as a planar sinusoidal waveform given in cartesian coordinates by
\begin{equation}
  y(x,t) = B \sin(2\pi x/\lambda-\omega t),
\end{equation}
where $B$ is the wave amplitude, $\omega$ the wave frequency and $\lambda$ its wavelength. The $x$ axis is the direction of propagation of the wave (see notation in  Fig.~\ref{wave}). As a result of the waving motion, the filament undergoes locomotion in the $-x$ direction. Since the filament is infinite, the swimming speed is expected to be steady. We non-dimensionalise length scales  by $\lambda/(2\pi)$ and times by $\omega^{-1}$, hence the waveform equation simplifies to $y =\epsilon\sin(x-t)$, where $\epsilon=2\pi  B/\lambda$, such that the non-dimensionalised wavespeed $V=\lambda\omega/(2\pi)=1$.   
The 
Newtonian drag coefficients are non-dimensionalised by the zero-shear viscosity.

For a slender  flagellum the force per unit length exerted by the fluid on the flagellum, $ \mathbf{f}$,  is quantified by the non-Newtonian resistive-force theory as 
\begin{equation}
 \mathbf{f}_{E/C} = -\left[R_{E/C}b\pa\uvec{t}\uvec{t}+R_{E/C}b\pe(\I-\uvec{t}\uvec{t})\right]\cdot\mathbf{u},
\end{equation}
where $\uvec{t}$ is the local tangent to the filament, $\mathbf{u}$ is the lab-frame velocity, and $R_{E}$ (resp~$R_{C}$) is the non-Newtonian correction factor for the Newtonian drag coefficients $(b\pa, b\pe)$ based on the experiments (resp.~on the Carreau model).  Classically, the non-dimensionalised velocity of each point along the flagellum  can  be written in the laboratory frame as \cite{LighthillLecture}
\begin{equation}\label{eq=u}
 \mathbf{u} = (1-U)\x - q\uvec{t}(s),
\end{equation}
where $q = \Lambda/\lambda >1$ is the ratio between the wavelength $\Lambda$ measured along the flagellum arc-length (s) and the wavelength $\lambda$ measured along the $x$ direction, $\x$ is the unit vector along the direction of the traveling wave, and $U<1$ the non-dimensionalised (unknown)  swimming speed. In order to determine the value of $U$ we enforce the free-swimming condition namely
\begin{equation}\label{eq:force}
 \int_0^{\Lambda}  \mathbf{f} \cdot\x ds=0,
\end{equation}
with the other components being zero by symmetry. The force density along $x$, $ f_x$, is given by 
\begin{equation}
 f_x = \mathbf{f}\cdot\x = -R_{E/C}b\pe\mathbf{u}\cdot\x + (R_{E/C}b\pe-R_{E/C}b\pa)(\mathbf{u}\cdot\uvec{t})(\uvec{t}\cdot\x).
\end{equation}
Using Eq.~\eqref{eq=u}, the above simplifies to
\begin{equation}\label{eqdfx}
 f_x = R_{E/C}b\pe(U-1)+R_{E/C}b\pa q\uvec{t}\cdot\x+(R_{E/C}b\pe-R_{E/C}b\pa)(1-U)(\uvec{t}\cdot\x)^2.
\end{equation}
Furthermore we define a dimensionless critical time to complete our non-dimensionalisation, where $\Gamma$ is rescaled to $\Gamma\omega$ (with identical notation retained for convenience)
such that the Carreau correction factor becomes
\begin{equation}
  R_C = \left[1+\left(\frac{\Gamma b\pe u_{\sh}}{2\sqrt{2}\pi a}\right)^2\right]^{\frac{n-1}{2}},
\end{equation}
where
\begin{equation}
 u_{\sh}^2 = (1-U)^2(\uvec{x}\cdot\uvec{n})^2+2\frac{b\pa^2}{b\pe^2}\left((1-U)\uvec{x}\cdot\uvec{t}-q\right)^2.
\end{equation}
Similarly upon non-dimensionalisation the experimental correction factor becomes,
\begin{equation}
 R_E =  \left[1+0.317\left(\frac{(1-n)\Gamma|\mathbf{u}|}{d_s}\right)^{0.692}\right]^{-1},
\end{equation}
where
\begin{equation}\label{eqUnd}
 |\mathbf{u}|^2 =  [(1-U) - q\uvec{t}\cdot\x]^2-(q\uvec{t}\cdot\y)^2.
\end{equation}
The unit tangent and normal to this wave are further given by
 $\uvec{t} = \left( \cos\theta , \sin\theta \right)  
$ and $
 \uvec{n}= \left( -\sin\theta ,   \cos\theta  \right) $ 
where $\theta$ is the angle between the swimming direction ($\x$) and the local tangent to the flagellum ($\uvec{t}$). 
We now have four dimensionless constants we are able to vary: 
$n$ and $\Gamma$ describing the fluid; $\alpha$ describing the aspect ratio of the flagellum; and $\epsilon$ describing the amplitude of the waveform.  
The variables $n$ and $\Gamma$ enter only through the correction factor whereas $\alpha$ and $\epsilon$ enter into both the correction factor and the Newtonian calculation through the Newtonian drag coefficients $b\pe$ and $b\pa$ which depend logarithmically on $\alpha$, and $\epsilon$ through the integral over the arc-length $s$.

Note that, as discussed above, the resistive-force theory description is only valid when the viscosity changes are local, hence in non-dimensional values the range of 
viable critical times is $\Gamma\lesssim0.6$, whose value is given for a typical spermatozoa flagella described in Section \ref{regime-valid}.

\subsection{Numerical implementation and  validation}
\begin{figure}[!]
\centering
\includegraphics[width=0.41\textwidth]{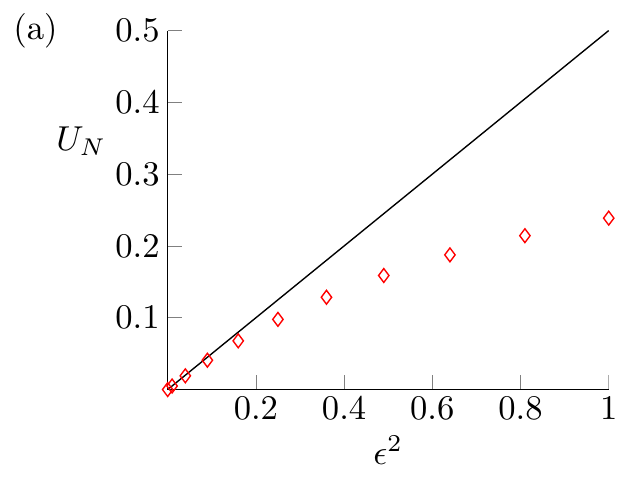}
\includegraphics[width=0.42\textwidth]{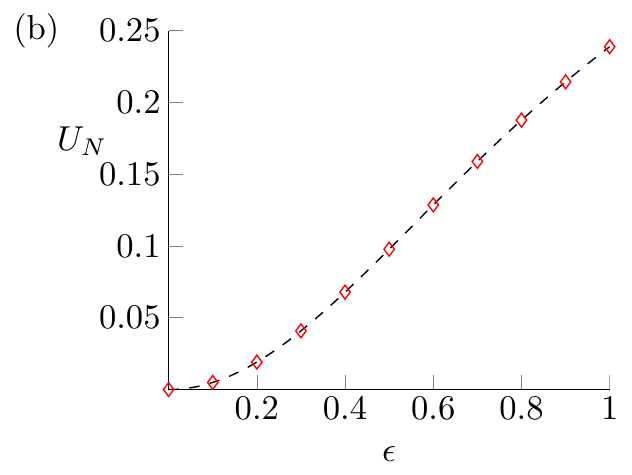}
\caption{The numerical Newtonian  results are compared to the analytic Newtonian results, (a) small-amplitude  analytical expansion (solid line, $\epsilon \ll 1$) compared to numerical results (red diamonds) and (b) full analytical result of Eq.~\eqref{eqUA}  (dashed line) compared to numerical results (red diamonds).}\label{fig:Unvalid}
\end{figure}

In order to validate our numerical implementation, we first address Newtonian swimming. 
We compare our numerical implementation of Eq.~\ref{eq:force} in the Newtonian limit to the analytic Newtonian result, 
namely \cite{LighthillLecture}
\begin{equation}
\label{eqUA}
U_N = \frac{\left(1-b\pa/b\pe\right)(1-\beta)}{1-(b\pa/b\pe-1)\beta},
\end{equation}
where 
\begin{equation}
 \beta = \frac{1}{\Lambda}\int^{\Lambda}_{0}\uvec{t}\cdot\x ds = \frac{1}{\Lambda}\int^{\Lambda}_{0}\cos^2\theta ds.
\end{equation}

At small amplitude $\epsilon \ll 1$, the  swimming speed limits to the asymptotic  result
\begin{equation}
 U_N = \frac{\epsilon^2(1-b\pa/b\pe)}{2b\pa/b\pe},
\end{equation}
which agree with our numerics when $\epsilon$ decreases as shown in Fig.~\ref{fig:Unvalid} (a).  A closer agreement can be found for all values of $\epsilon$ by comparing our full numerical results to the swimming speed given in Eq.~\eqref{eqUA}, where $\beta$ is evaluated numerically. The comparison is show  in 
Fig.~\ref{fig:Unvalid} (b) and  therefore validates our numerical implementation of the free-swimming problem.

\subsection{Non-Newtonian locomotion}
\begin{figure}[!]
\includegraphics[width=0.42\textwidth]{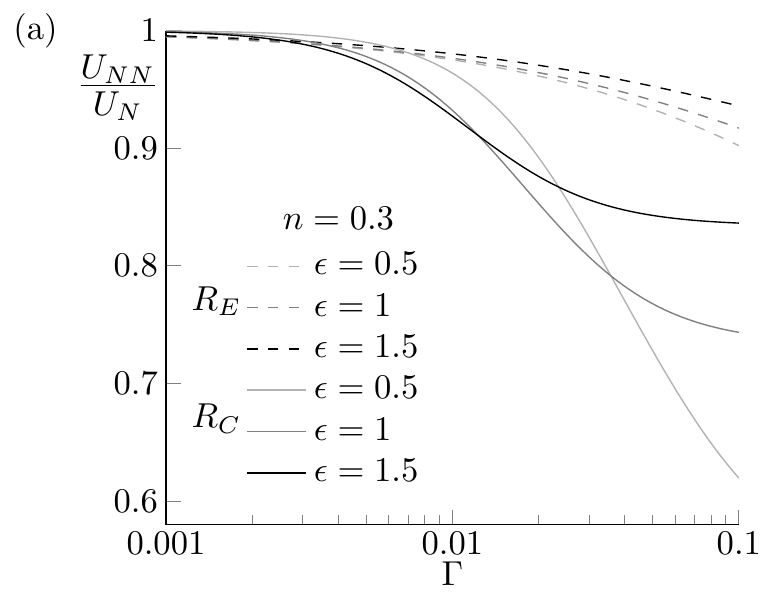}
\includegraphics[width=0.42\textwidth]{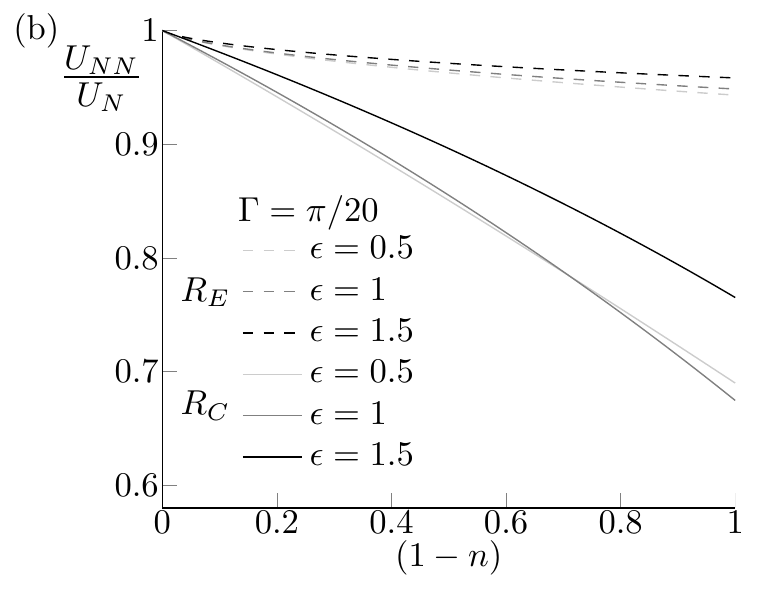}
\caption{Ratio between the non-Newtonian  and Newtonian swimming speeds, $U_{NN}/U_N$, as a function of the properties of the fluid. 
(a) speed ratio with fixed $n=0.3$ for a range of amplitudes $\epsilon$ plotted against the critical fluid time $\Gamma$.
(b) speed ratio with fixed value of $\Gamma=\pi/20$ 
for a range of wave amplitudes $\epsilon$ plotted against the power index of the fluid, $n$.  The swimming speed is always reduced in a shear-thinning fluid. 
}\label{fig:nnfluidprop}
\end{figure}

We now follow a similar numerical approach to tackle the non-Newtonian problem, however as the integral depends on the velocity $U$, we must solve this iteratively, taking the Newtonian solution as the initial value from which we iterate. The main results are shown in Fig.~\ref{fig:nnfluidprop} where we plot the ratio between the non-Newtonian swimming speed of the waving flagellum ($U_{NN}$) and the Newtonian one ($U_N$) as a function of the critical time of the fluid $\Gamma$ (a) and as a function of the power index $n$ (b). Results for the two models are superimposed: the Carreau approach is plotted in solid lines while the experimentally-based model is show in dashed line, each for a few different values of the wave amplitude $\epsilon$.  

While the Carreau and experimental models do not agree quantitatively, they both provide a similar physical picture. Under this modelling approach, swimming of a waving flagellum is always slower in a non-Newtonian fluid  than in a Newtonian fluid, and all the more that the critical time  $\Gamma$ increases Fig.~\ref{fig:nnfluidprop} (a) or that the power index $n$ decreases ~\ref{fig:nnfluidprop} (b). Both  illustrate that, for a fixed geometry, the greater the non-Newtonian effects in the fluid the slower the resulting swimming speed.

\begin{figure}[!]
\includegraphics[width=0.42\textwidth]{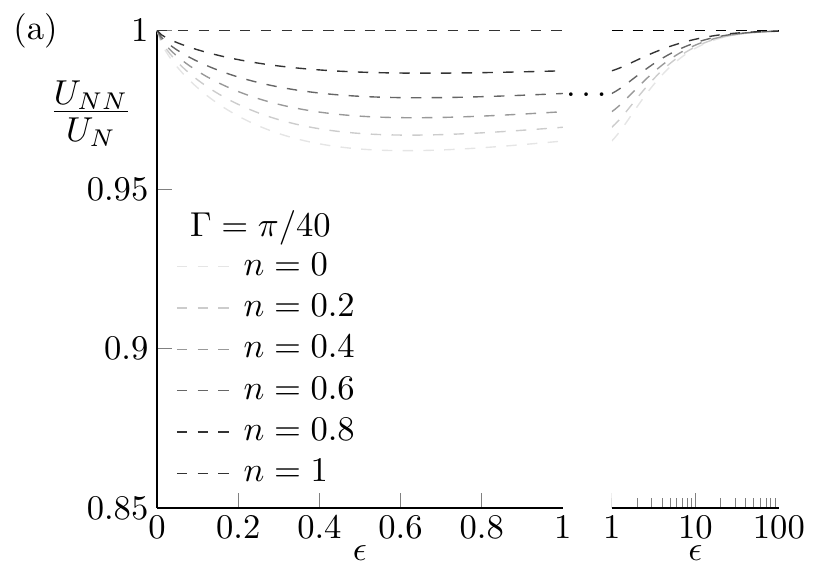}
 \includegraphics[width=0.42\textwidth]{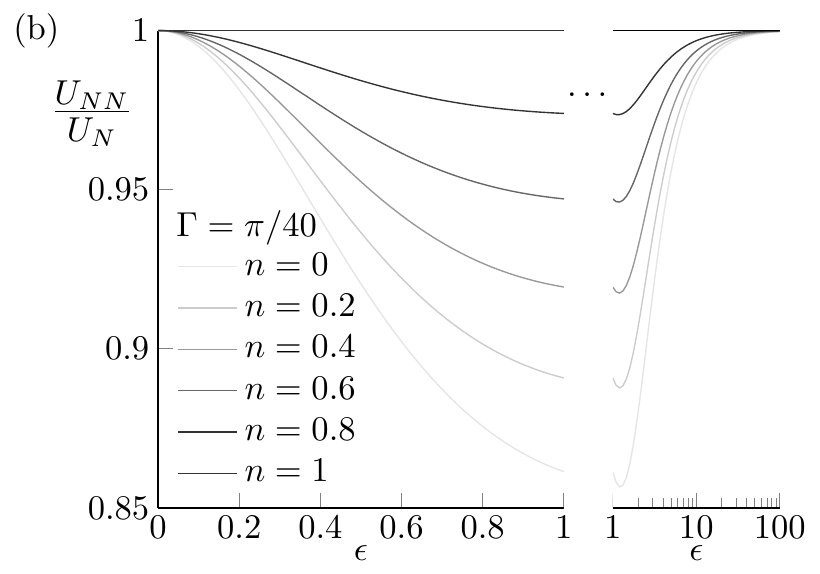}
\caption{Ratio of the non-Newtonian to Newtonian swimming speeds $U_{NN}/U_N$ for fixed critical time  $\Gamma=\pi/40$, for a range of power indices $n$ as a function of the flagellum amplitude $\epsilon$, (a) experimental model ($R_E$) and (b) Carreau model ($R_C$).}\label{fig:nnwaveprop}
\end{figure}

We further see in Fig.~\ref{fig:nnfluidprop} that the results for different wave amplitudes, $\epsilon$,  do not collapse onto the same curve; non-Newtonian swimming has therefore a different $\epsilon$ dependence from Newtonian swimming.  To address this we plot the ratio of the swimming speeds against $\epsilon$ for a fixed $\Gamma$ value and a range of power indices in Fig.~\ref{fig:nnwaveprop}. For both models, we observe  a non-monotonic dependence of the swimming speed ratio on the wave amplitude $\epsilon$. 
The swimming speed always  deceases with $\epsilon$ for small amplitudes, reaches  a minimum 
for the Carreau correction factor and experimental correction factor when $\epsilon\approx1.2$ and $\epsilon\approx0.6$ respectively (the precise value depends in fact on the fluid properties). Finally, at large amplitudes, the swimming speed ratio asymptotes again to $U_{NN}/U_N\approx 1$.

Unlike the Newtonian result, both non-Newtonian swimming speeds depend on the rod shape \emph{i.e.}~the value of $\alpha$, as the correction factors depend on $\alpha$ in such a way that it cannot be factored out of the integral equation.
As with $\Gamma$, both the experimental  and Carreau model swimming speeds decrease monotonically with increasing $\alpha$, showing that we would expect  fatter swimmers to be hindered more by the shear-thinning fluid. These results are not shown here as they are  qualitatively similar to  the dependence on $\Gamma$ in Fig.~\ref{fig:nnfluidprop} (a).

\subsection{Asymptotic results}

In order to understand further the systematic decrease in swimming speed in a Newtonian  results we turn to considering  the impact of a small amount of non-Newtonian rheology to the fluid. The transition between a Newtonian fluid and a shear-thinning fluid occurs for fluids with a finite critical time and with a  shear-thinning index below unity.  
In the experimental correction factor, both terms $(1-n)$ and $\Gamma$ appear as a single factor $\chi_E \equiv [(1-n)\Gamma]^{0.692}$. 
In the Carreau model we must assume that $(1-n)$ is small to ensure that the non-Newtonian effects are small. Since $(1-n)$ only appears as a power in the correction factor $R_C$, the lowest order non-zero term is
$\chi_C \equiv (1-n)\Gamma^2$. 

\begin{figure}[!]
\includegraphics[width=0.4\textwidth]{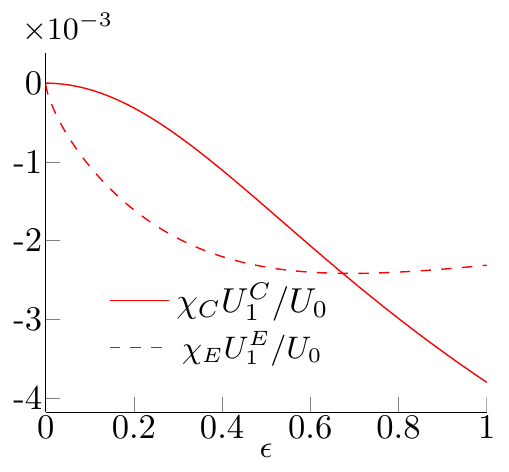}
\caption{First-order corrections to the Newtonian velocity scaled by the Newtonian speed, $U_1/U_0$, for the Carreau model (solid line) and the experimental model (dashed line) as a function of the wave amplitude  $\epsilon$. The correction is always negative indicating a decrease in the swimming speed.}\label{fig:expchi}
\end{figure}

We can then expand mathematically both empirical models about the small parameters $\chi_E$ and $\chi_C$ respectively, leading to swimming with speeds written at first order as  $U^{(E)}_{NN}\approx U_0+\chi_E U^{(E)}_1$ and $U^{(C)}_{NN}\approx U_0+\chi_C U^{(C)}_1$ respectively. At zeroth order for both models the Newtonian result is recovered such that $U_0 = U_N$. In order to expand the velocity-dependent drag correction factors we must insert $U^{(E)}_{NN}$ and $U^{(C)}_{NN}$ into their respective correction factors $R_E$ and $R_C$, such that
$R_E\approx 1+\chi_ER_{E(1)}$ and $R_C\approx1+\chi_CR_{C(1)}$ where
\begin{equation}
 R_{E(1)} = -0.317\left(\frac{|\mathbf{u}_0|}{d_s}\right)^{0.692},
\end{equation}
and
\begin{equation}
 R_{C(1)} = -\frac{1}{2}\left(\frac{b\pe u_{{\sh}_0}}{2\sqrt{2}\pi a}\right)^2,
\end{equation}
are the first order correction factors, in which
\begin{equation}
 |\mathbf{u}_0| = \sqrt{(1+q^2)+U_0^2+2q(U_0-1)\cos\theta-2U_0},
 \end{equation}
 and
 \begin{equation}
 u_{{\sh}_0} = \sqrt{(1-U_0)^2(\uvec{x}\cdot\uvec{n})^2+2\frac{b\pa^2}{b\pe^2}\left((1-U_0)\uvec{x}\cdot\uvec{t}-q\right)^2},
 \end{equation}
are the leading order rod section velocity and shear rate velocity respectively.
After expansion, the first-order  experimental  and Carreau swimming speeds are obtained to be 
\begin{equation}\label{asymRE}
 U_1^{(E)} = -0.317\left(\frac{1}{d_s}\right)^{0.692}\frac{\displaystyle\intarc |\mathbf{u}_0|^{0.692}P(\theta)ds}{\displaystyle b\pe\Lambda-\beta(b\pe-b\pa)\Lambda},
\end{equation}
and
\begin{equation}\label{asymRC}
 U_1^{(C)} = -\frac{1}{2}\left(\frac{b\pe}{2\sqrt{2}\pi a}\right)^2\frac{\displaystyle\intarc u_{{\sh}_0}^2P(\theta)ds}{b\pe\Lambda-\beta(b\pe-b\pa)\Lambda},
\end{equation}
respectively, where
\begin{equation}
 P(\theta) = (1-U_0)b\pe-b\pa\cos\theta-(1-U_0)(b\pe-b\pa)\cos^2\theta,
\end{equation}
where $\theta$ is implicitly a function of the arc-length $s$. 
The right-hand side of both Eqs.~\eqref{asymRE} and \eqref{asymRC} only depend on the Newtonian results and can thus be easily evaluated numerically. The  results for both $ U_1^{(E)} $ and $ U_1^{(C)} $ are   shown in Fig.~\ref{fig:expchi}. Both the experimental and Carreau first-order swimming speeds are  negative for all values of the wave amplitude indicating a decrease in the swimming speed with non-Newtonian effects in agreement with our full numerics. 

\subsection{Physical interpretation}

In order to gain some fundamental understanding on the origin of the observed systematic reduction in swimming speed, we take a closer look at the distribution of shear-rates along the waving flagellum.
As the non-Newtonian equations are too nonlinear to glean physical insight, we consider the  Newtonian shear-rates to inform our understanding of the system. We use the shear-rates calculated for our Carreau correction factor with the knowledge that larger shear-rates will lead to reduced drag force for both our correction factors. 

We consider separately the ``thrust'' problem, where $U=0$ and a net force is a applied on the fluid,  and the ``drag'' problem, where $V=0$ and the flagellum is dragged passively through the fluid.  In the thrust problem, denoted `$th$',  the magnitude of non-dimensionalised shear-rate velocity is given by
\begin{equation}
u_{\sh_{th}} = \sqrt{\sin^2\theta+2\frac{b\pa}{b\pe}\left(\cos\theta-q\right)^2},
\end{equation}
leading to a shear-rate along the flagellum of
\begin{equation}
 |\sh_{th}| = \frac{b\pe u_{\sh_{th}}}{2\sqrt{2}\pi a},
\end{equation}
 where shear-rates are non-dimensionalised by $\omega$.
In the  drag problem, denoted by `$dr$',  since the non-dimensionalised relevant shear-rate velocity is given by
\begin{equation}
u_{\sh_{dr}} = U\sqrt{\sin^2\theta+2\frac{b\pa}{b\pe}\cos^2\theta},
\end{equation}
the shear-rate is then given by
\begin{equation}
 |\sh_{dr}| = \frac{b\pe u_{\sh_{dr}}}{2\sqrt{2}\pi a}\cdot
\end{equation}

\begin{figure}[!]
\includegraphics[width=0.4\textwidth]{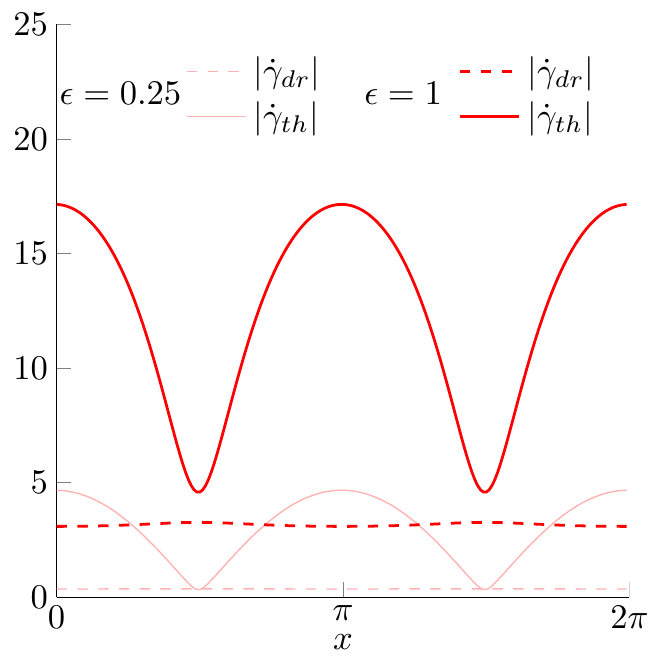}
\caption{Dimensionless shear-rates along a waving flagellum  as a function of the dimensionless position $x$  along the flagellum for the drag problem (dashed lines) and the thrust problem (solid lines), shown for a small amplitude ($\epsilon=0.25$) and  large amplitude ($\epsilon=1$) swimmer.}
\label{figwavetd}
\end{figure}

The distribution of  shear-rates for both  thrust  and  drag  problems is shown in 
Fig.~\ref{figwavetd} for two wave amplitudes (small in thin line and large in thick line). Note that the drag problem is computed for $U=U_N$ i.e.~we are comparing the shear-rates for the flow around the flagellum for the two problems which, on average, induce equal and opposite forces during the swimming motion. What is apparent from these results is that the shear-rates in the thrust problem are systematically larger than in the drag problem essentially everywhere along the waving flagellum. In a shear-thinning fluid, the higher the shear-rate the lower the viscosity. Since for swimming thrust and drag have to balance, we see that if the swimming speed was kept constant, forces would not balance and there would be more  drag than thrust. The vale of the swimming speed has thus to decrease in order to compensate for it. 
 
An alternative way to interpret this result is to consider the case of waving at small amplitude $\epsilon$.  In that limit, the shear-rates for the thrust  and drag problems are given by
\begin{equation}
 |\sh_{th}|\approx\frac{b\pe\epsilon|\cos(x-t)|}{2\sqrt{2}\pi} + O(\epsilon^2),
\end{equation}
and,
\begin{equation}
 |\sh_{dr}|\approx \frac{b\pa\epsilon^2}{2\sqrt{2}\pi}\frac{\left(1-b\pa/b\pe\right)}{b\pa/b\pe} + O(\epsilon^4),
\end{equation}
respectively. As the  shear-rate in the drag problem is a factor  $\epsilon$ smaller than the one due to  thrust generation, we obtain a relatively larger reduction in thrust, and thus a reduction in the swimming speed. Fundamentally, the difference in shear-rate scaling between thrust  ($\sim \epsilon$) and drag ($\sim \epsilon^2$)  arises from the fact that in the thrust-producing waving motion,    
only a small subset of the  periodic up-and-down motion in the direction perpendicular to the swimming direction   is rectified  to produce useful work for swimming.

\section{Comparison with \emph{C. elegans} experiments}
\label{Snematode}

In order to  demonstrate the relevance of our empirical model we now compare our simulations results to the experimental results of Gagnon et al.~\cite{Arratia2014}, where the swimming motion of the small nematode \emph{Caenorhabditis elegans} (\emph{C. elegans}) was studied in both Newtonian and  shear-thinning  fluids.  The shear-thinning fluids in their study are composed of Xanthan gum solutions, shown by rheological measurements to be inelastic and well described by the Carreau model, with greater shear-thinning obtained for larger Xanthan gum concentrations. \emph{C. elegans} are then immersed in the different fluids within an acrylic chamber of diameter 2~cm and thickness 1~mm.  The organisms are approximately 1~mm in length and 80~$\mu$m in diameter. Through body tracking, the swimming speed, frequency, wavespeed and amplitude of  the waving nematode is measured in each of the different fluids.  These are then plotted against an effective viscosity  $\eta_{eff}$, defined as the average viscosity over the shear-rates experienced by the swimmer, $U/L\leq\sh\leq 2\omega B/d$, and ranging in this series of experiments from 6~mPa~s to 200~mPa~s. The results for shear-thinning fluids are then compared to Newtonian fluids with similar viscosities to the effective viscosities of the shear-thinning fluids.

\begin{figure}[!]
\includegraphics[width=0.49\textwidth]{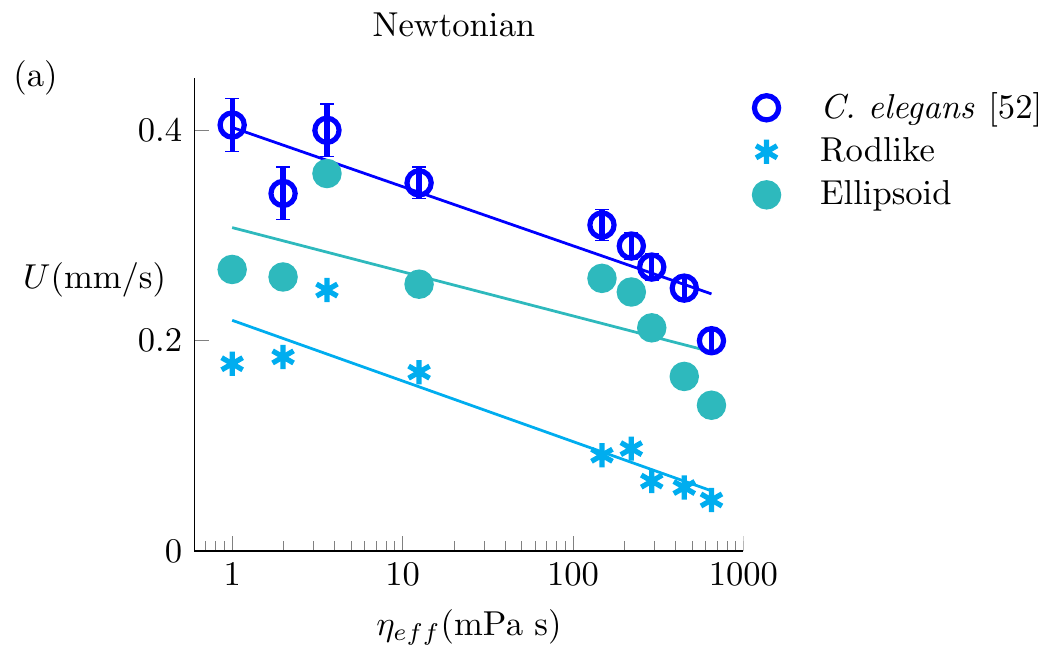}
\includegraphics[width=0.49\textwidth]{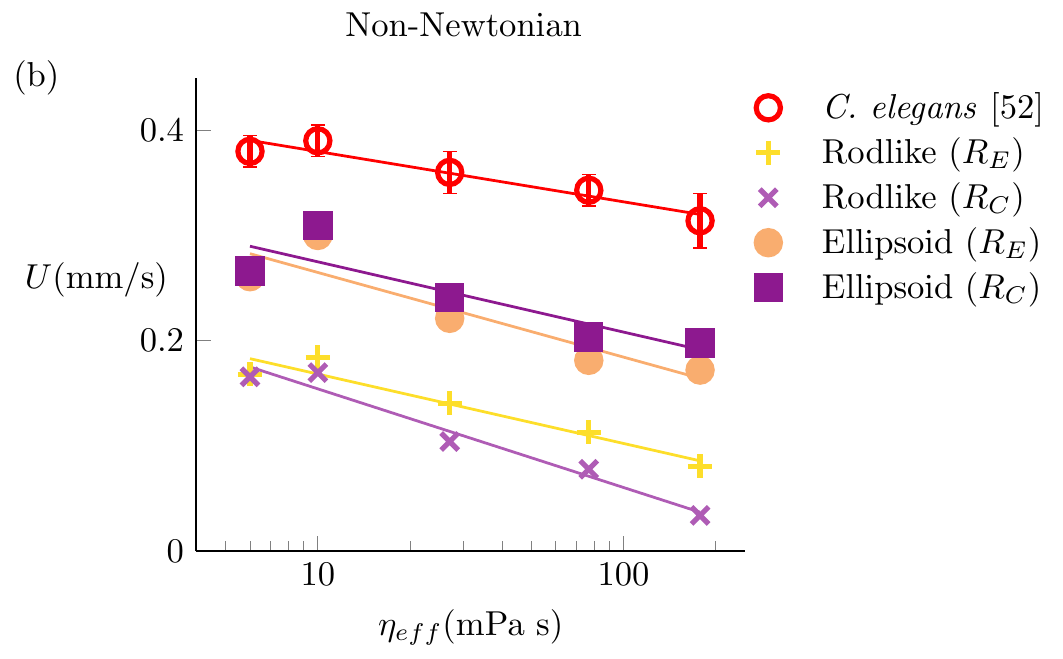}
\caption{Experimental results from Ref. \cite{Arratia2014} (open symbols with error bars) plotted together with our simulation results (line or filled symbols).   Newtonian swimming speeds are shown in (a), and non-Newtonian swimming speeds in (b).  The simulation results are represented by line symbols for thin rod results and filled symbols for fat rod results.  The lines in each of the different colours are straight lines of best fit to their matching colour symbol. Each shear-thinning simulation data point shares fluid and swimmer properties with the  red experimental shear-thinning swimming speed at that particular effective viscosity.}
\label{expcomp}
\end{figure}

Using the experimental data from Ref.~\cite{Arratia2014}, the waveform and wavespeed of each swimmer in the different fluids is known as well as the fluid properties, and thus a direct comparison with our model results can be made with no fitting parameters. This comparison is presented in Fig.~\ref{expcomp} where we plot the  dimensional swimming speed, $U$,  against the effective viscosity, $\eta_{eff}$.  Each  simulation data point shares fluid and swimmer properties with the   experimental shear-thinning swimming speed at that particular effective viscosity. Experimental results are shown with open symbols   (each data point represents the mean and standard error of approximately 15 experiments \cite{Arratia2014}) while the results of our models are shown with line symbols (rodlike filaments) and  filled symbols (ellipsoidal filaments). Specifically we use line symbols (crosses and stars) to plot results of our modelling approach using $\ell=0.09\lambda$, as described in Section \ref{Srods} (both experimental and Carreau model), here the Newtonian drag ratio is give by $b\pe/b\pa\approx1.1$. As wavelength is larger than the length of the nematode, $\ell$ is too short instead we choose the body length $\ell_{nema}$ and note that its shape is more accurately  described by that of a prolate ellipsoid of aspect ratio $\alpha=d_{nema}/l_{nema}\approx0.08$, where $d_{nema}$ is the nematode diameter. 
 Using the drag coefficients described in Ref.~\cite{BookMircoHydro} for prolate spheroids, we find that the dimensions of \emph{C. elegans}  correspond to  a drag ratio $b\pe/b\pa\approx1.5$,  which is within the range of drag ratios calculated by Ref. \cite{Berman2013} for biologically relevant swimming. The corresponding results are show with filled circles in 
 Fig.~\ref{expcomp} for the experimental empirical model and the Carreau empirical model.

While some discrepancies exist, we see that both sets of numerical simulations share the qualitative features of the experimental results, which are greater than all modelling in both Newtonian and non-Newtonian fluids. Our empirical non-Newtonian resistive-force theory is thus able to capture the main physical features of swimming in a shear-thinning fluid. Quantitative differences are expected to arise from multiple sources. First, our model is confined to local effects of the thinning fluid  whereas we estimate the fluid to be thinned over 100 nematode radii in the experiments. This would lead to  soft confinement effects, and an increase of the swimming above that shown by the thick filament model in Fig.~\ref{expcomp} (b) \cite{Man2015,li15}.  Furthermore our simulations do not include end effects, which are predicted to increase swimming speeds in non-Newtonian fluids \cite{Becca2014, Shelley2013}.  

Wall effects in the confined experimental setup are also expected to play a role whereas our model considers swimming in  an infinite fluid. Despite these possible sources of discrepancies, our simple empirical model is able to capture the main physical features of waving locomotion in a shear-thinning fluid.

\section{Conclusion}

Flagella waving in fluids are expected to be  subject to  two types of physical changes when the fluid is no longer Newtonian but is shear-thinning \cite{gomez16}. The first one, local, is due to the decrease of the fluid forces resulting from a decrease of the fluid viscosity. The second, nonlocal,  results from overall changes to the flow field in the fluid, and is similar to enhanced in swimming under soft confinement  \cite{Man2015,li15}. In this paper we proposed an empirical model to quantify the first of these effects by replacing the classical Newtonian drag coefficients with velocity-dependent shear-thinning drag coefficients based on experimental results or empirical modelling.  We illustrated our new models by calculating the swimming speed of an infinite planar wave swimmer with a slender flagellum, for a range of shear-thinning fluid parameters,  and apply our results   to a set of experimental results on \emph{C. elegans}  taking into account the ellipsoidal shape of the nematode \cite{Arratia2014}. 

The main limitation to our model, beyond the fact that it is clearly not derived from first principles and is thus empirical, is the small range of relaxation times (or actuation frequencies) where our model is viable. Indeed, as with Newtonian resistive-force theory, we must ensure that the flow induced by the moving portion of a filament is local otherwise interactions between different sections of the filament are required. In a shear-thinning fluid this means that the flow outside the ``region of influence'' of size $\ell$ needs to be Newtonian to allow fluid stresses to be  determined solely by the local kinematics.   This imposes thus a limit on the range of shear-rates between the critical shear-rate and the largest shear-rate experienced by the flagella.  Typical shear-rates generated by spermatozoa, cilia and \emph{C. elegans} are in the range $10^1-10^3\,\rm{s}^{-1}$, and typical critical shear-rates of mucus are on the order $|\sh_{C}|\sim10^{-3}\,\rm{s}^{-1}$ \cite{velezcordero13},  and soil $|\sh_{C}|\sim10^{-1}\,\rm{s}^{-1}$ \cite{SoilThesis}. Hence the typical distance $r$ away from a particular location along the flagellum where  the fluid is Newtonian is on the order $r/a\sim10^2$ for \emph{C. elegans} in soil and $r/a \sim10^4-10^5$ for cilia and spermatozoa in mucus. Meaning the fluid around biological organisms is already heavily sheared by the motion at a length $\ell$ before the swimmer, and therefore the requirement $\Gamma\nu\leq0.09$ is likely to not be reached in vivo. 

While the work presented here focused on planar waving motion it could  can be adapted to helical propulsion of bacteria. In that case, and unlike for planar swimming, the presence of a head is crucial to balance hydrodynamic moments  \cite{ChildressBook}.  The force integral over the rigid helical flagella must match the force generated by the head, and similarly for the torque, then the rotation rate and the swimming speed can be obtained. In order to describe the force and torque on the head both the rotational and translational the drag coefficient of the head would be required.  If the head is rod-shaped then the translational drag coefficients are as described in this paper, however for spherical (e.g.~\emph{coccus}) or more complex head shapes knowledge of new drag coefficients would be required, obtained experimentally or numerically. 

Resistive-force theory has also been used to tackle  large variety of problems in the biophysics of swimming cells. With our modelling approach,  these results could then be extended to more complex fluids.  Problems which could be tackled include the polymorphic transitions of bacteria flagella \cite{Vogel2012}, bundling of flagella \cite{Powers2002}, swimming non-flagellated bacteria \cite{Yang2009}, the generation of waving modes in passive \cite{Passov2012, Yu2006} and active filaments \cite{Camalet1998}, and the motion of filaments in external flows  \cite{Marcos2009, Kim2014}.

\section*{Acknowledgements}
This work was funded in part by the  the EPSRC (E.E.R.) and by the 
 European Union through a Marie Curie CIG grant (E.L.) and an ERC consolidator grant (E.L.).

\bibliographystyle{unsrt}
\bibliography{refsNNRFT}

\end{document}